\documentclass[aps,prl,twocolumn,groupedaddress]{revtex4}
\usepackage{graphicx}
\usepackage{psfrag}
\usepackage{epsfig}

\newcommand{\be}{\begin{equation}}
\newcommand{\ee}{\end{equation}}

\newcommand{\beq}{\begin{equation}}

\newcommand{\eeq}{\end{equation}}

\newcommand{\bea}{\begin{eqnarray}}
\newcommand{\eea}{\end{eqnarray}}
\newcommand{\ba}{\begin{eqnarray}}
\newcommand{\ea}{\end{eqnarray}}

\begin{document}

\title{
Recovering General Relativity from massive gravity
}
\author{E.~Babichev}
\author{C.~Deffayet} 
\author{R.~Ziour}

\affiliation{AstroParticule \& Cosmologie,
UMR 7164-CNRS, Universit\'e Denis Diderot-Paris 7,
CEA, Observatoire de Paris,
10 rue Alice Domon et L\'eonie
Duquet, F-75205 Paris Cedex 13, France}

\begin{abstract}
We obtain static, spherically symmetric, and asymptotically flat numerical solutions of massive gravity with a source.
Those solutions show, for the first time explicitly, a recovery of the Schwarzschild solution of General Relativity via the so-called Vainshtein mechanism.

\end{abstract}

\date{\today}


\maketitle

It is now well established that the expansion of the Universe is accelerating. 
This can be explained by a non vanishing cosmological constant, raising various questions at the fundamental level, 
 or by a ubiquitous, but mysterious, dark energy.
  Instead of  introducing such new dark components in the Universe, so far only detected via their gravitational effects, a logical alternative is to consider possible large distance modifications of gravity. 
 Historically, the introduction of the cosmological constant has been linked by Einstein himself to the possibility of modifying gravity at large distance via a Yukawa decay \cite{SCHU}, i.e. to give a mass to the graviton. Interestingly, ``massive gravity'' has recently attracted a lot of interest in this context via constructions relying on models with extra-dimensions like the Dvali-Gabadadze-Porrati (DGP) model \cite{DGP}. This model, where gravity is mediated via a resonance of massive gravitons, indeed produces a late time acceleration of the expansion of the Universe via a modification of gravity at cosmological distances \cite{ced}. The consistency of the DGP model has been questioned 
 and variants have been proposed \cite{DEGRAV}. On the other hand, the way well tested predictions of General Relativity (GR) are recovered in models with massive gravitons, is highly non trivial. Indeed, a consistent free massive graviton is uniquely defined by the  Pauli-Fierz theory \cite{PF},  which is well know to lead to different physical predictions from those of GR, irrespectively of how small the graviton mass is. 
This is the celebrated van Dam-Veltman-Zakharov (vDVZ) discontinuity  \cite{VDVZ}. This, based on observations such as light bending around the Sun, would be enough to reject any theory of massive gravity provided it is well approximated by the Pauli-Fierz theory. However, soon after it was discovered, it was pointed out by A.~Vainshtein that the vDVZ discontinuity could be cured in a suitable interacting extension of Pauli-Fierz theory \cite{Vainshtein:1972sx}. The purpose of this letter is to show that this ``Vainshtein mechanism'' is indeed working for static spherically symmetric solutions appropriate to describe the metric outside spherically symmetric sources. Although a cosmological version of the Vainshtein recovery of GR was shown to be present for exact solutions of the DGP model \cite{Deffayet:2001uk}, nothing analogous was known so far for static solutions; e.g. static spherically symmetric solutions of DGP model are only known locally in some approximation scheme \cite{SPHEDGP}. Our results are not only of interest for the specific interacting massive gravity studied here, but have also implications for more sophisticated models such as the DGP model or its variants \cite{DGP,DEGRAV}, the ``degravitation'' models \cite{DEGRAVbis}, the Galileon and k-Mouflage models \cite{GALIK}, as well as possibly the Ho\v{r}ava-Lifshitz theory \cite{Horava:2009uw} (see e.g. \cite{Mukohyama:2009tp}).

The theory we will consider
 has one dynamical metric $g_{\mu \nu}$ non derivatively coupled to a background non dynamical flat metric $f_{\mu \nu}$ 
via a scalar interaction term ${\cal V}_{int}[f,g]$. It is defined by the following action
\begin{equation}
\label{action} S=  \frac{M_P^2}{2}\int d^4 x \sqrt{-g} \left( R_{(g)} -\frac{1}{4} m^2 {\cal V}_{int}[f,g] \right),
\end{equation}
where $R_{(g)}$ is the Ricci scalar built with $g_{\mu \nu}$.
The interaction  ${\cal V}_{int}[f,g]$ is chosen such that, when expanded at quadratic order around a flat background for $g_{\mu \nu}$, one recovers the Pauli-Fierz mass term with a graviton mass given by $m$. There are infinitely many possible such choices \cite{Damour:2002ws}; we shall investigate here one of the simplest forms,
 also considered in \cite{Arkani-Hamed:2002sp,Damour:2002gp}, given by
\begin{equation}
{\cal V}_{int} = ( g_{\mu \nu} - f_{\mu \nu})(g_{\sigma\tau} - f_{\sigma \tau})\left(g^{\mu\sigma}g^{\nu\tau}-g^{\mu\nu}g^{\sigma\tau}\right)\label{S3}.
\end{equation}
We also consider that matter is minimally coupled to the dynamical metric $g_{\mu \nu}$. 
The equations of motion, obtained varying action (\ref{action}) with respect to the metric $g_{\mu \nu}$, read
\be \label{EQMot}
M_P^2 G_{\mu \nu} =T_{\mu \nu}+ T^{(g)}_{\mu \nu},
\ee
where $G_{\mu\nu}$ denotes
the Einstein tensor computed with the metric $g_{\mu \nu}$,
$T_{\mu \nu}$ is the energy momentum tensor of matter fields, and
$T^{(g)}_{\mu \nu}$ is an effective energy momentum tensor coming from the variation with respect to the metric $g_{\mu \nu}$ of the interaction term $ \sqrt{-g} \times {\cal V}_{int} $.

An ansatz for $g_{\mu \nu}$ appropriate to describe static spherically symmetric solutions reads  
\be \label{gnl}
g_{\mu \nu}dx^\mu dx^\nu = -e^{\nu(R)} dt^2 + e^{\lambda(R)} dR^2 + R^2 d\Omega^2 ,
\ee
where $R$ is parameterizing the distance to the source. With such a choice, the solution for $g_{\mu \nu}$ can easily be compared to the Schwarzschild solution of GR  given by Eq. (\ref{gnl}) with $\nu = \nu_{GR} \equiv \ln(1-R_S/R)$ and $\lambda = \lambda_{GR} \equiv - \nu_{GR}$, where $R_S$ is the Schwarzschild radius of the source. For simplicity, the metric $f_{\mu \nu}$ is taken to cover a Minkowski space-time. However one can see that when one chooses the ansatz (\ref{gnl}) for $g_{\mu \nu}$, one should keep  
one unknown function $\mu(R)$ in $f_{\mu \nu}$ in order for the field equations not to be overconstrained. Hence, 
we take $f_{\mu \nu}$ in the non canonical flat-space form \cite{Vainshtein:1972sx,Salam:1976as,Damour:2002gp}
\ba
f_{\mu \nu}dx^\mu dx^\nu &=& - dt^2 + \left(1-\frac{R \mu '(R)}{2}\right)^2 e^{-\mu(R)} dR^2 \nonumber \\
&& + e^{-\mu(R)}R^2 d\Omega^2,
\label{fmu}
\ea
where here and henceforth 
a prime denotes a derivative with respect to $R$.
Note that $\mu$ cannot be measured directly because it only appears in $f_{\mu \nu}$, and hence, in the following we will  focus on the physical components 
$\{\nu, \lambda\}$. To describe a static spherically symmetric source, we consider a matter energy momentum tensor in the perfect fluid form reading $T_{\mu}^{\;\nu}=\text{diag} (-\rho, P,P,P)$,
where the pressure $P$ and energy density $\rho$ depend only on $R$. We also assume that this energy momentum tensor vanishes outside the radius $R_\odot$ of the source.
For such a perfect fluid and the above ansatz (\ref{gnl}), the conservation equation reads 
\be \label{cons}
P'= -\frac{\nu'}{2}\left(P+\rho\right).
\ee
We look for asymptotically flat and everywhere non singular solutions. In particular, we impose boundary conditions such that $\mu$, and $\nu$ vanish at infinity as well as $\lambda$ and $\mu'$ in  $R=0$.

If one linearizes the vacuum equations of motion, one finds that $\nu$ and $\lambda$ are respectively given by 
\bea
\nu_L &=& -{\cal C} e^{-mR}/R, \label{nuL}\\
\lambda_L &=&  {\cal C} (1+mR) e^{-mR}/(2R), \label{lL}
\ea where ${\cal C}$ is some integration constant. Notice that for $R \ll m^{-1}$ one gets that 
 $\nu_L \sim -2 \lambda_L$
 irrespectively of the graviton mass. 
  This is in contrast with the corresponding relation for GR stating that at distances larger than $R_{\odot}$ one has $\nu_{GR} = -\lambda_{GR}$,
 and is just another way to phrase the vDVZ discontinuity. 
The constant ${\cal C}$ should be fixed by including a source.

Following Vainshtein, one can then show that the solution $\{\nu_{L}, \lambda_{L}\}$ obtained by linearization ceases to be valid at a distance smaller than the Vainshtein radius, $R_V$, given by $R_V = \left(m^{-4}R_S\right)^{1/5}$. For a graviton with cosmologically large Compton length $m^{-1}$, the Vainshtein radius of the Sun is much larger than the solar system size impeding the use of the linearized solution to compare with solar system observables. Vainshtein argued that, below $R_V$, a recovery of the standard solution of GR was possible. Indeed, he found a local expansion of the functions $\{\lambda, \nu\}$ around their GR expressions $\{\lambda_{GR}, \nu_{GR}\}$ that  stays valid at small $m$ and for distances $R \ll R_V$ (an expansion with similar properties was also found for $\mu$). For the theory (\ref{action}-\ref{S3}) this expansion reads (for $R \gg R_S$) 
\ba
\nu- \nu_{GR} &\sim&   (mR)^2 \sqrt{(8 R_S)/(81 R)} + {\cal O}(m^4), \label{V1}\\
\lambda - \lambda_{GR} &\sim&  - (mR)^2 \sqrt{ (2 R_S)/( 9 R)} + {\cal O}(m^4). \label{V2}
\ea
In contrast, the linearized expressions $\{\lambda_{L}, \nu_{L}\}$ are nowhere valid when $m$ goes to zero, because $R_V$ is infinite in this limit. However it was pointed out by Boulware and Deser \cite{Boulware:1973my} that  there was no warranty that the two expansions found by Vainshtein (for $R \gg R_V$ and $R\ll R_V$) could be joined into a non singular solution of the field equations. This question remained unanswered for a while and it was more recently investigated in Ref. \cite{Damour:2002gp} which concluded, on the basis of a numerical investigation of the field equations, that no such solution existed and hence that the Vainshtein mechanism, i.e. the non perturbative recovery of solution of massless gravity, does fail in the massive gravities defined by the class of action (\ref{action}), 
 and in particular for theory (\ref{action}-\ref{S3}). Here we reexamine this question in the light of a recent work \cite{US} where we considered  static spherically symmetric solutions of the  field equations (\ref{EQMot}) in the the so-called decoupling limit (DL).
Note that the theories in the class (\ref{action}) suffer from the Boulware-Deser instability \cite{Boulware:1973my} associated with a mode that starts propagating at the nonlinear level. This is probably enough to make those theories unrealistic, but is not really worrisome for the present work, since we simply view the theory (\ref{action}-\ref{S3}) as a tool to investigate the validity of the Vainshtein mechanism, and do not advocate for a realistic application of this particular theory.

The DL enables to focus on the nonlinearities responsible for the crossover behaviour arising at the Vainshtein radius $R_V$ and is obtained by sending together $M_P$ and $m^{-1}$ to $\infty$ while keeping the energy scale $\Lambda = (m^4 M_P)^{1/5}$, as well as $R_S \times M_P$, fixed. The scale $\Lambda$ is associated with the strongest self-interaction in the scalar sector of the model \cite{Arkani-Hamed:2002sp}. The DL should 
provide a good description of the solutions (provided those exist) in the range of distances $R_S \ll R \ll m^{-1}$. Notice that this range of distance covers in particular the crossover distance $R_V$ which stays constant in the DL. In \cite{US} we obtained numerical solutions of the DL field equations which are showing a transition between a $R \ll R_V$ regime recovering linearized GR behaviour and a $R \gg R_V$ regime, where the leading behaviour of the solutions is given by Eqs. (\ref{nuL}-\ref{lL}) where one sends $m \times R$ to zero. Such solutions were shown to be everywhere non singular and to exist only for a class of theories of the kind (\ref{action}) including the one defined by Eq. (\ref{S3}).
Note that, in the DL 
 the nonlinearities important in GR at the Schwarzschild radius are absent as well as the terms responsible for the Yukawa decay at large distances.
A crucial property of the DL, is that, in vacuum, we found infinitely many solutions decreasing at infinity. This was shown both by numerical integration 
and by using Ecalle's theory of resurgent functions  and, in particular,  allows one to accommodate the presence of a non vanishing source.

\begin{figure}[t]
  \begin{center}
    \includegraphics[height=\columnwidth,angle=-90]{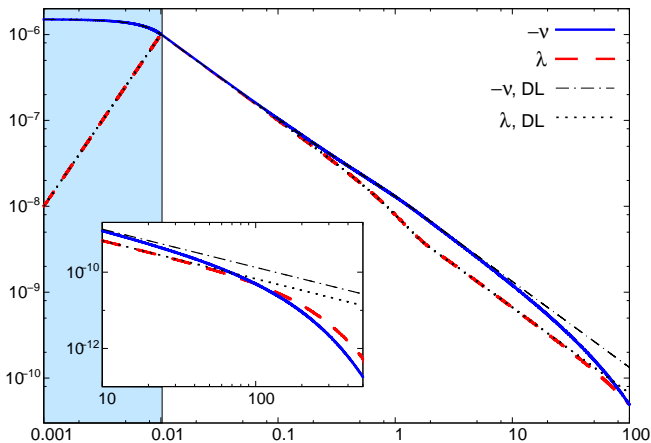}
    \caption{\label{fig1} Plot of the metric functions $-\nu$ and $\lambda$ vs. $R/R_V$, in the full nonlinear system and the decoupling limit (DL), with a star of radius $R_{\odot} =10^{-2} R_V$
and $ m \times R_V = 10^{-2}$. For $R \ll R_V$, the numerical solution is close to the GR solution (where in particular $\nu \sim -\lambda \sim- R_S/R$ for $R > R_{\odot}$).
For $R \gg R_V$, the solution enters a linear regime. Between $R_V$ and $m^{-1}$, where the DL is still a good approximation, one has $\nu \sim - 2 \lambda\sim -4/3 \times R_S/R$. At distances larger than $m^{-1}$ the metric functions decay {\it \`a la} Yukawa as appearing more clearly in the insert. The latter shows the same solution but for larger values of $R/R_V$, and in the range of distance plotted there, the numerical solutions are indistinguishable from the analytic solutions of the linearized field equations Eqs.~(\ref{nuL}-\ref{lL}).}
  \end{center}
\end{figure}

The existence of non singular solutions in the DL  is
not enough to conclude on the validity of the Vainshtein mechanism in the full theory. First because those solutions are only covering at best a limited range of distances $R_S \ll R \ll m^{-1}$ (this interval being sent to $[0,+\infty]$ taking the DL) and second because it could be that nonlinearities present in the original field equations, but left aside by the DL, destabilize  the DL solutions
enough to make them become singular.

In this letter we report results of the numerical integration of the full nonlinear field equations  (\ref{EQMot})-(\ref{cons}) for the theory (\ref{action}-\ref{S3}) and metric ans\"atze (\ref{gnl}-\ref{fmu}).
Those equations, whose explicit (but rather long) form can be found in Ref.\cite{Damour:2002gp}, 
reduce to a set of quasilinear ordinary differential equations of first order in $\lambda$ and $\nu$ and of second order in $\mu$. We included a static spherical source of radius $R_\odot$ and constant density $\rho_{\odot}$. Hence, defining the Schwarzschild radius $R_S$ of the source as $R_S \equiv M_P^{-2} R_{\odot}^3 \rho_{\odot}/3$, there are only two dimensionless parameters in the model which can be taken to be $R_{\odot}/R_V$ and $a \equiv m \times R_V$. We were able to find everywhere non singular asymptotically flat solutions, $\{\nu,\lambda, \mu, P\}$, by solving numerically the field equations, using both shooting and relaxation methods. To our knowledge, the existence of such solutions was unknown and shows for the first time explicitly that the Vainshtein mechanism can work in a theory of massive gravity. The shooting approach, which is based on a direct Runge-Kutta integration, is quite difficult to implement since the initial conditions need to be extremely finely tuned in order to obtain non singular solutions. This was already the case in the DL \cite{US}
and could explain the findings of Ref. \cite{Damour:2002gp}. We have been able to integrate the system of equations outward from a point deep inside the source up to a point further than the Vainshtein radius (typically $R\sim 3 R_{V}$). We have also been able to integrate inward starting from a point far from the source (\textit{e.g.} $R\sim 3 R_{V}$) up to the core of the source. In fact the integration of the vacuum equations shows that, as in the DL, one is able to obtain infinitely many decaying solution by slight changes of the boundary condition. In all the integrations, we have made an extensive use of local  analytic series expansions of the solution in order to fix the initial conditions as precisely as possible.  Even in cases where we could check that those expansions were only asymptotic expansions (i.e. that the series were not convergent), they were very useful to set with enough precision the boundary data of the numerical integrations.
\begin{figure}[t]
  \begin{center}
    \includegraphics[height=\columnwidth,angle=-90]{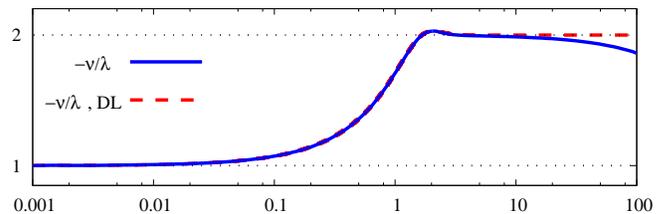}
    \caption{\label{fig2} Plot of the ratio of $\nu$ to $-\lambda$ vs. $R/R_V$, with a star of radius $R_{\odot}/R_V =10^{-3}$
and $ m \times R_V = 10^{-3}$. This shows the transition, at the Vainshtein radius, between a GR regime where $\nu \sim -  \lambda$ to a  regime where $\nu \sim - 2 \lambda$.
At larger distance, the solution features the expected Yukawa cutoff.}
  \end{center}
\end{figure}Details of the numerical integration will be reported elsewhere \cite{USFUT}. In order to be able to obtain a wide enough range of integration (typically, $R$ runs between $R =10^{-5} R_V$ and $R \sim3 R_V$) the required relative precision on the initial conditions is usually of order $10^{-8}$ and often even more stringent. These results, obtained through a Runge-Kutta integration, have been confirmed by the use of a relaxation method. 
In particular the relaxation approach allowed us to solve the system on a wide range of parameters and distances (typically for $R$ between $0$ and $\sim 100 \; R_{V}$). Both methods perfectly agree, demonstrating the robustness of our numerical investigation. Our main finding is that for $a \in [10^{-3},0.6]$ and sources that are not too compact (i.e. $R_{\odot} \gtrsim 5 R_{S}$), a non singular solution can be found, which is asymptotically flat and has the right boundary conditions at the origin $R=0$. It features a recovery of GR at distances $R \ll R_V$, where we also checked numerically that the first correction to the GR behaviour was given by the form given in Eqs. (\ref{V1}-\ref{V2}). It is also well approximated by the solution obtained in the DL in the expected range of distances $R_S \ll R \ll m^{-1}$. In particular, at distances above $R_V$ (but below $m^{-1}$) the behaviour of the solution is given by the  regime 
\be \nu \sim - 2 \lambda \sim - 4/3 \times R_S/R , \ee  
as obtained from Eqs. (\ref{nuL}-\ref{lL}). 
Note that the constant ${\cal C}$ is found to be $ \sim 4/3 \times R_S$ by the numerical integration. This value of ${\cal C}$ agrees with that  obtained  in the linearized theory 
for a non relativistic point-like source placed at the origin, and can be explained by the presence of a scalar polarization also responsible for the vDVZ discontinuity. 
At distances above $m^{-1}$ the solution features the exponential Yukawa falloff. We checked numerically that this falloff agrees with the one given in Eqs. (\ref{nuL}-\ref{lL}) as well as the fact 
that our solutions were continuously connected to flat space-time, when the mass of the source is decreased either by keeping the density fixed but decreasing the radius or by keeping the radius fixed but decreasing the density. Those results are depicted in Figs. \ref{fig1}-\ref{fig3}.

\begin{figure}[t!]
  \begin{center}
    \includegraphics[height=\columnwidth,angle=-90]{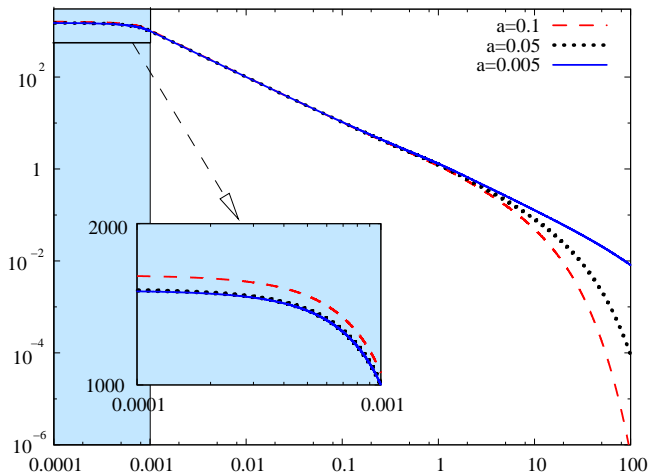}
    \caption{\label{fig3} Plot of $-\nu \times a^{-4}$ vs. $R/R_V$ (the $a^{-4}$ factor is included for convenience such that, in the decoupling limit (DL), all plotted theories would exactly coincide) for three different values of $a \equiv  m \times R_V$ and source of radius $R_{\odot} = 10^{-3} R_V$. The solution with $a=0.005$ corresponds to a spherical source of size and density close to that of the Milky-Way and a graviton Compton length of the order of the Hubble radius. In the range of distances plotted, this solution (with $a=0.005$) is well approximated by the DL solution. The insert shows a zoom on small distances. There, the behaviour of our solutions agrees with the one of GR, and for $a=0.1$, a value which belongs to a parameter range investigated in Ref. \cite{Damour:2002gp}, one can see that the solution departs from the DL, emphasizing the role of the non-linearities of GR which become important.}   
\end{center}
\end{figure}

To summarize, we found static, spherically symmetric, and asymptotically flat numerical solutions of massive gravity showing, for the first time explicitly, a recovery of the usual Schwarzschild solution of General Relativity via the so-called Vainshtein mechanism. Some issues deserve more investigations. These include the classical stability of our solutions as well as  an understanding of highly relativistic objects and Black Holes.

\begin{acknowledgments}
We thank T.~Damour for interesting discussions and suggestions. The work of E.~B. was supported by the EU FP6 Marie Curie Research and Training Network ``UniverseNet'' (MRTN-CT-2006-035863). 
\end{acknowledgments}

\end{document}